\pdfoutput=1
\documentclass{article}%
\usepackage{amssymb}
\usepackage{amsfonts}
\usepackage{amsmath}
\usepackage{graphicx}%
\providecommand{\U}[1]{\protect\rule{.1in}{.1in}}
\newtheorem{theorem}{Theorem}

\newtheorem{lemma}[theorem]{Lemma}

\newtheorem{proposition}[theorem]{Proposition}

\begin{document}

\title{A.Cormack's last inversion formula and a FBP reconstruction}
\author{V. P. Palamodov}
\date{28.07.2014}
\maketitle

\textbf{Abstract }

A reconstruction of a function from integrals over the family of confocal
paraboloids in $\mathbb{R}^{n}$ is given by a FBP formula.

\section{Introduction}

Cormack proposed a clever method for determination\ of distribution of a
scattered material in a space from data of scattered light filtered on travel
times \cite{Co5}. He wrote in the introduction to \cite{Co5}: "Suppose that
waves travel in this space with a speed $v,$ and that the space contains a
distribution of scattering material with a density $f$ which is assumed to be
smooth and either rapidly decreasing or of compact support... An impulsive
plane wave front with a normal $\omega$ reaches $0$ at $t=0.$ As result of the
well-known focusing property of paraboloids the scattering waves arriving at
$0$ at time $t=2p/v$ will all have originated on the paraboloid defined by
$r\left(  1+\left\langle \xi,\omega\right\rangle \right)  =2p..."$ The
mathematical problem is to reconstruct a function in a plane from data of
integrals $\mathrm{R}f$ over a family of confocal parabolas or confocal
rotation paraboloids. "The problem of determining $f$ from $\mathrm{R}f$ for a
family of curves in $\mathbb{R}^{2}$ was discussed in \cite{Co1}, \cite{Co2},
\ \cite{Co3} and in \cite{Co3} it was shown that the parabolic case could be
reduced to the ordinary Radon transform by a coordinate transformation which
does not have an analog in $\mathbb{R}^{3}.$ The case of $\mathbb{R}^{n}$ was
considered in \cite{Co4}, in which $f$ was considered to be expanded in
spherical harmonics; some properties of the harmonic components were given,
and a solution to the resulting integral equation was given. This solution,
however, had a very awkward form and as a result it was not investigated in
detail. Given in \cite{Co5} is a simpler solution for the case of
$\mathbb{R}^{3}...$" This solution is%

\begin{equation}
r^{1/2}f\left(  x\right)  =-\frac{1}{\left(  4\pi\right)  ^{2}}\int
_{\mathrm{S}^{2}}\left(  \frac{\partial}{\partial r}r\right)  ^{2}\left.
\frac{g\left(  p,\omega\right)  }{p^{3/2}}\right\vert _{p=r\left(
1+\left\langle \xi,\omega\right\rangle \right)  /2}\mathrm{\Omega,}\label{2}%
\end{equation}
where $r=\left\vert x\right\vert ,\ \xi=x/\left\vert x\right\vert $,
$\mathrm{\Omega}$ is the area form in $\mathrm{S}^{2}$ and $g$ is the surface
integral of $f.$ 

Our goal is a reconstruction by means of an integral transform of FBP type to
avoid divergent integrals.

\section{Confocal paraboloids and integrals}

Confocal parabolas

\begin{center}
\fbox{\includegraphics[
natheight=2.352300in,
natwidth=3.312200in,
height=2.3523in,
width=3.3122in
]{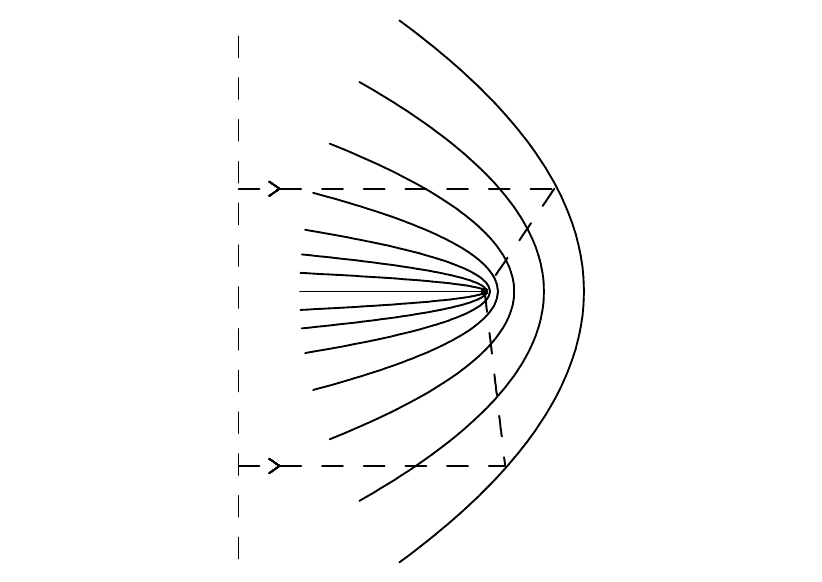}}\newline
\end{center}

Reflecting paraboloid

\begin{center}
\fbox{\includegraphics[
natheight=3.000000in,
natwidth=4.499600in,
height=3in,
width=4.4996in
]{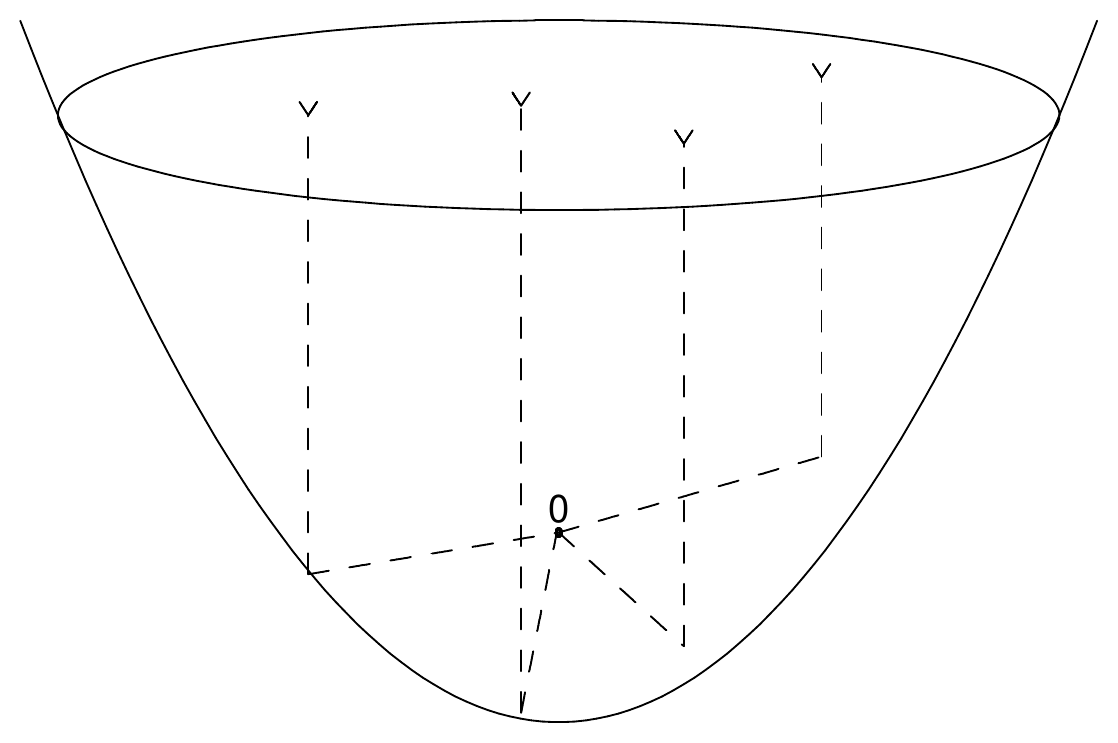} }\newline
\end{center}

The family of paraboloids $Z\left(  p,\omega\right)  \ $in an Euclidean space
$E^{n}$ with the focus at the origin can be given by the
equation$\ r-\left\langle x,\omega\right\rangle =2p\ $where $p>0$ and
$\omega\in\mathrm{S}^{n-1}$ are parameters. For a bounded function $f$ in
$E^{n}$ with compact support we consider the integrals over paraboloids%
\begin{align}
\mathrm{R}f\left(  p,\omega\right)   &  =\int_{Z\left(  p,\omega\right)
}f\mathrm{d}S,\nonumber\\
\mathrm{M}f\left(  p,\omega\right)   &  =\lim_{\varepsilon\rightarrow0}%
\frac{1}{\varepsilon}\int_{p\leq r-\left\langle x,\omega\right\rangle \leq
p+\varepsilon}f\mathrm{d}x=\int_{Z\left(  p,\omega\right)  }\frac
{f\mathrm{d}S}{\left\vert \nabla\theta\left(  x,\omega\right)  \right\vert
},\label{3}%
\end{align}
where $\mathrm{d}S$ is the Euclidean surface density and $\mathrm{M}f$ is
equal to the integral of $f\mathrm{d}x$ between to close paraboloids. The
function can be reconstructed from $\mathrm{R}f$ and from $\mathrm{M}f$
according to the following

\begin{theorem}
\label{C}For an arbitrary function $f\in C^{2}\left(  E^{2}\right)  $ with
compact support,$\ $a reconstruction is given by
\end{theorem}%

\begin{align}
f\left(  x\right)   &  =-\frac{1}{4\pi^{2}r^{1/2}}\int_{\mathrm{S}^{1}}%
\int_{\mathbb{R}}\frac{\partial}{\partial p}\left(  p^{1/2}\mathrm{R}f\left(
p,\omega\right)  \right)  \frac{\mathrm{d}p}{\theta\left(  x,\omega\right)
-p}\mathrm{\Omega}\label{10}\\
&  =-\frac{1}{4\pi^{2}r}\int_{\mathrm{S}^{1}}\int_{\mathbb{R}}\frac{\partial
}{\partial p}\left(  p\mathrm{M}f\left(  p,\omega\right)  \right)
\frac{\mathrm{d}p}{\theta\left(  x,\omega\right)  -p}\mathrm{\Omega,\ }x\in
E^{2}\backslash0.\nonumber
\end{align}
For $f\in C^{2}\left(  E^{3}\right)  ,$ we have
\begin{align}
f\left(  x\right)   &  =-\frac{1}{8\pi^{2}r^{1/2}}\int_{\mathrm{S}^{2}}%
\frac{\partial^{2}}{\partial p^{2}}\left.  p^{1/2}\mathrm{R}f\left(
p,\omega\right)  \right\vert _{p=\theta\left(  x,\omega\right)  }%
\mathrm{\Omega}\label{11}\\
&  =-\frac{1}{8\pi^{2}r}\int_{\mathrm{S}^{2}}\frac{\partial^{2}}{\partial
p^{2}}\left.  p\mathrm{M}f\left(  p,\omega\right)  \right\vert _{p=\theta
\left(  x,\omega\right)  }\mathrm{\Omega,\ }x\in E^{3}\backslash0.\nonumber
\end{align}

\textbf{Remark.} The first formula (\ref{11}) reminds Cormack's reconstruction
(\ref{1}) where apparently $g=\mathrm{R}f.$ However the integral $\mathrm{M}f$
is a correct model for the "scattered light filtrated on travel times". It is
different from the Euclidean surface integral $\mathrm{R}f$ since the
dominator in (\ref{3}) is not a constant.

\textbf{Proof. }We assume at the beginning that $n\geq2$ is arbitrary. The
family of rotation paraboloids is\ generated by the function $\Phi\left(
x,p,\omega\right)  =\theta\left(  x,\omega\right)  -p,$ $\theta\left(
x,\omega\right)  =\left(  r-\left\langle x,\omega\right\rangle \right)  /2$
defined in the manifold $X\times\Sigma,\ X=E^{n}\backslash0,\ \Sigma
=\mathbb{R}_{+}\times\mathrm{S}^{n-1}.$ The set $X$ is supplied with the
Euclidean volume form $\mathrm{d}x$ and the manifold $\Sigma$ with the volume
form $\mathrm{d}p\wedge\mathrm{\Omega.}$ The map $\pi_{X}:Z\rightarrow X$ is
proper where $Z=\Phi^{-1}\left(  0\right)  $ and the function $\Phi$ but
satisfies\ the conditions: $\pi_{X}$ has rank $n$ and the\textit{\ }%
map\textit{\ }$\pi^{\ast}$\textit{\ }is a\textit{\ }bijection see \S 3. We
have for any $n\geq2$%
\begin{align*}
\nabla_{x}\theta &  =\frac{\xi-\omega}{2},\ \text{where }\xi=\frac{x}%
{r},\ \text{and\ }\left\vert \nabla_{x}\theta\right\vert =\theta^{1/2}%
r^{-1/2},\\
2^{n}\left(  n-1\right)  !J\left(  x,\omega\right)   &  =\nabla\theta
\wedge\left(  \mathrm{d}_{\omega}\nabla\theta\right)  ^{\wedge n-1}%
=\left\langle \xi-\omega,\omega\right\rangle \ \omega\wedge\left(
\mathrm{d}\omega\right)  ^{\wedge n-1}=\left\langle \xi-\omega,\omega
\right\rangle \mathrm{\Omega.}%
\end{align*}
It follows that $\det J\left(  x,\omega\right)  \neq0$ except for the
submanifold $V\doteqdot\left\{  \omega=\xi\right\}  $ in $X\times\Sigma.$
However $\pi^{\ast}$ is a bijection since $\left\langle \xi-\omega
,\omega\right\rangle \geq0.$ Choose a weight function $b\left(  x,\omega
\right)  =\left\vert \nabla_{x}\theta\left(  x,\omega\right)  \right\vert
^{2}$ and consider the weighted Funk transform%
\begin{equation}
\mathrm{M}_{b}f\left(  p,\omega\right)  =\int_{\theta\left(  x,\omega\right)
=p}f\left(  x\right)  \left\vert \nabla_{x}\theta\left(  x,\omega\right)
\right\vert \mathrm{d}S=p^{1/2}\mathrm{R}\left(  r^{-1/2}f\right)
=p\mathrm{M}\left(  r^{-1}f\right)  ,\label{13}%
\end{equation}
where $\mathrm{R}$ is the Euclidean integral transform as above. We want to
invert $\mathrm{M}_{b}$ by means of a filtered back projection. Take a number
$\varepsilon>0$ and consider the $\varepsilon$-neighborhood $V\left(
\varepsilon\right)  $ . The generating function $\Phi$ satisfies (\textbf{i})
for the map $\pi^{\ast}$ restricted to the set $Z\backslash V\left(
\varepsilon\right)  \times\mathbb{R}_{+}.$ Let $b\left(  \varepsilon\right)  $
be a function in $X\times\Sigma$ that vanishes in $V\left(  \varepsilon
\right)  $ and coincides with $b$ otherwise. Take $\beta=1$ as the second
weight functions and apply Theorem 5\S 3 to $\Phi$ and $Z\backslash V\left(
\varepsilon\right)  .\ $Note that equation $\mathrm{d}_{\omega}\theta\left(
x,\omega\right)  =-x/2$ implies (\textbf{ii}). This yields%
\[
D_{b\left(  \varepsilon\right)  }f=\mathrm{NM}_{b\left(  \varepsilon\right)
}f+\Theta_{b\left(  \varepsilon\right)  }f,
\]
where $\mathrm{N}=\mathrm{N}_{\beta}$ and $\Theta_{b\left(  \varepsilon
\right)  }$ is the operator with the kernel%
\[
\Theta_{b\left(  \varepsilon\right)  }\left(  x,y\right)  =\operatorname{Re}%
i^{n}\int_{\mathrm{S}^{2}}\frac{b\left(  \varepsilon\right)  \left(
y,\omega\right)  \mathrm{\Omega}}{\left(  \theta\left(  x,\omega\right)
-\theta\left(  y,\omega\right)  -i0\right)  ^{n}}.
\]
By (\ref{6})%
\begin{equation}
D_{b\left(  \varepsilon\right)  }\left(  x\right)  =\frac{1}{\left\vert
\mathrm{S}^{n-1}\right\vert }\int_{\mathrm{S}^{n-1}}\frac{b\left(
\varepsilon\right)  \mathrm{\Omega}}{\left\vert \nabla_{x}\theta\right\vert
^{n}}.\label{8}%
\end{equation}

\begin{lemma}
\label{L1}We have $D_{b\left(  \varepsilon\right)  }\left(  x\right)
\rightarrow D_{b}\left(  x\right)  $ for $x\neq0$ and $\Theta_{b\left(
\varepsilon\right)  }\left(  x,y\right)  \rightarrow0$ for arbitrary $x\neq
y\in E^{n}\backslash0$ as $\varepsilon\rightarrow0$ where
\end{lemma}%

\begin{equation}
D_{b}\left(  x\right)  =\frac{1}{\left\vert \mathrm{S}^{n-1}\right\vert }%
\int_{\mathrm{S}^{n-1}}\frac{b\mathrm{\Omega}}{\left\vert \nabla_{x}%
\theta\right\vert ^{n}}=\frac{1}{\left\vert \mathrm{S}^{n-1}\right\vert }%
\int_{\mathrm{S}^{n-1}}\frac{\mathrm{\Omega}}{\left\vert \nabla_{x}%
\theta\right\vert ^{n-2}}=2^{n-2}\label{9}%
\end{equation}
for any $n\geq2.$

\textbf{Proof} of Lemma \ref{L1}. We have $\left\vert \nabla_{x}%
\theta\right\vert ^{2}=r^{-1}\theta,$ hence $\left\vert \nabla\theta\left(
x,\omega\right)  \right\vert ^{-1}\approx\sin\gamma/2$ where $\gamma$ is the
angle between $x$ and $\omega.$ Therefore integral (\ref{9}) converges to
(\ref{8}). This proves the first statement. To check the second statement we
note that
\[
\theta\left(  x,\omega\right)  -\theta\left(  y,\omega\right)  =\frac{1}%
{2}\left(  \left\vert x\right\vert -\left\vert y\right\vert -\left\langle
x-y,\omega\right\rangle \right)
\]
\ is a linear function of $\omega$ for arbitrary $x\neq y.$ This function has
a zero $\omega\in\mathrm{S}^{n-1}$ since $\left\vert \left\vert x\right\vert
-\left\vert y\right\vert \right\vert \leq\left\vert x-y\right\vert .$ By
\cite{Pa} it follows that $\Theta_{b}\left(  x,y\right)  =0$ since
$\theta\left(  x,\omega\right)  -\theta\left(  y,\omega\right)  $ is a linear
function of $\omega$ and $\deg b+n-1=n.$ Now it is sufficient to check that
$\Theta_{b\left(  \varepsilon\right)  }\rightarrow\Theta_{b}$ as
$\varepsilon\rightarrow0.$ For this we show that the dominator $\theta\left(
x,\omega\right)  -\theta\left(  y,\omega\right)  $ does not vanish at the
point $\omega=\xi\doteqdot x/r.$ We have%
\[
\theta\left(  x,\xi\right)  -\theta\left(  y,\xi\right)  =\frac{-\left\vert
x\right\vert \left\vert y\right\vert +\left\langle y,x\right\rangle
}{2\left\vert x\right\vert }<0
\]
since $y\neq x$ and $x\neq0,y\neq0.$ This completes the proof.
$\blacktriangleright$

By Lemma\ \ref{L1} we can take the limit in (\ref{12}) and obtain the
equation$\ D_{b}f=\mathrm{NM}_{b}f.$ Due to (\ref{13}) we can replace the
integrals \textrm{M}$_{b}f$ by data of $\mathrm{M}\left(  r^{-1}f\right)  $
and or by data of $\mathrm{R}\left(  r^{-1/2}f\right)  .$ By substitution
$f=r\tilde{f}$ or $f=r^{1/2}\tilde{f}$ we finally come up to the equations
(\ref{10}) and (\ref{11}). This completes the proof of Theorem \ref{C}.
$\blacktriangleright$

\begin{proposition}
Integrals (\ref{10}) and (\ref{11}) have the classical sense for sufficiently
smooth functions $f.$
\end{proposition}

\textbf{Proof. }For $n=2,$ we have
\[
\mathrm{d}Z=\left(  \mathrm{d}r^{2}+r^{2}\mathrm{d}\gamma^{2}\right)
^{1/2}=\frac{\mathrm{d}r}{\cos\gamma/2}=\frac{\mathrm{d}r}{\left(
1-p/r\right)  ^{1/2}}%
\]
and%
\begin{align*}
p^{1/2}\mathrm{R}\left(  r^{1/2}f\right)  \left(  p,\omega\right)   &
=p^{1/2}\int_{\sin^{2}\left(  \gamma/2\right)  =p/r}r^{1/2}f\left(
r,\gamma\right)  \mathrm{d}Z\\
&  =2\int_{p}^{\infty}f\left(  r,\gamma\right)  \frac{r\mathrm{d}r}{\left(
r/p-1\right)  ^{1/2}}=2\int_{1}^{\infty}f\left(  sp,\gamma\right)
\frac{s\mathrm{d}s}{\left(  s-1\right)  ^{1/2}}.
\end{align*}
The right hand side has continuous second derivative with respect to $p$ and
$\omega$ if $f\in C^{2}\left(  E^{2}\right)  .$

\begin{lemma}
\label{L2}For any $f\in C^{4}\left(  E^{3}\right)  ,$ we have
\[
\mathrm{M}f\left(  p,\omega\right)  =p^{-1/2}\mathrm{R}\left(  r^{1/2}%
f\right)  \left(  p,\omega\right)  =g_{0}\left(  \omega\right)  +pg_{1}\left(
p,\omega\right)
\]
as $p\rightarrow0$ where $g_{0}$ and$\ g_{1}$ are $C^{2}$-continuous functions.
\end{lemma}

Theorem \ref{C} will follow since $\mathrm{M}f=p^{-1/2}\mathrm{R}\left(
r^{1/2}f\right)  $ and%
\[
\frac{\partial^{2}}{\partial p^{2}}\left(  p\mathrm{M}f\left(  p,\omega
\right)  \right)  =\frac{\partial^{2}}{\partial p^{2}}\left(  p^{1/2}%
\mathrm{R}\left(  r^{1/2}f\right)  \left(  p,\omega\right)  \right)
=2g_{1}\left(  p,\omega\right)  +o\left(  1\right)  .
\]

\textbf{Proof.} Variables $r,\ \gamma$ and $\varphi$ are spherical coordinates
in $\mathbb{R}^{3}$ where $\gamma,\ 0\leq\gamma<\pi$ is the spherical distance
between points $\xi,\omega\in\mathrm{S}^{2}$ and $\varphi$ is rotation angle
about $\omega$ and $Z\left(  p,\omega\right)  $. Equation$\ p=\left(
x-\left\langle x,\omega\right\rangle \right)  /2=r\sin^{2}\left(
\gamma/2\right)  \ $defines the surface $Z\left(  p,\omega\right)  $ and we
have%
\[
\mathrm{d}Z=\left(  \mathrm{d}r^{2}+r^{2}\mathrm{d}\gamma^{2}\right)
^{1/2}r\sin\gamma\mathrm{d}\varphi
\]
for the Euclidean area $\mathrm{d}Z$ of $Z\left(  p,\omega\right)  $. We have%
\[
\left(  \mathrm{d}r^{2}+r^{2}\mathrm{d}\gamma^{2}\right)  ^{1/2}%
=\frac{\mathrm{d}r}{\cos\gamma/2},\ \mathrm{d}Z=2\left(  pr\right)
^{1/2}\mathrm{d}r\mathrm{d}\varphi.
\]
This yields%
\[
p^{-1/2}\mathrm{R}\left(  r^{1/2}f\right)  \left(  p,\omega\right)
=p^{-1/2}\int_{\sin^{2}\left(  \gamma/2\right)  =p/r}r^{1/2}f\left(
r,\gamma,\varphi\right)  \mathrm{d}Z=2\int_{p}^{\infty}\int_{0}^{2\pi
}rf\left(  r,\gamma,\varphi\right)  \mathrm{d}\varphi\mathrm{d}r.
\]
For a plane $P$ and a function $\phi\in C^{4}\left(  P\right)  ,\ $the circle
integral can be represented in the form%
\[
\int_{\left\vert y\right\vert =s}\phi\left(  y\right)  \mathrm{d}\varphi
=2\pi\phi\left(  0\right)  +\tau\left(  s\right)  s^{2}%
\]
with a remainder $\tau\in C^{2}.$ Applying this equation to $f$ in a plane $P$
orthogonal to $\omega$ for $s=r\sin\gamma$ we obtain
\[
\int_{0}^{2\pi}f\left(  r,\gamma,\varphi\right)  \mathrm{d}\varphi=2\pi
f\left(  r,0.\varphi\right)  +\left(  r\sin\gamma\right)  ^{2}h\left(
r,\gamma,\varphi\right)  ,\ h\in C^{2}.
\]
We have $\left(  r\sin\gamma\right)  ^{2}=4p\left(  r-p\right)  $ and
$f\left(  r,0,\varphi\right)  =f\left(  r\omega\right)  $ which implies%
\[
p^{-1/2}\mathrm{R}\left(  r^{1/2}f\right)  \left(  p,\omega\right)  =4\pi
\int_{p}^{\infty}rf\left(  r\omega\right)  \mathrm{d}r+8p\int_{p}^{\infty}%
\tau\left(  r,\gamma,\omega\right)  \left(  r-p\right)  r\mathrm{d}r.
\]
Now Lemma \ref{L2} follows for the functions
\[
g_{0}\left(  \omega\right)  =4\pi\int_{p}^{\infty}rf\left(  r\omega\right)
\mathrm{d}r,\ g_{1}\left(  p,\omega\right)  =8\int_{p}^{\infty}\tau\left(
r,\gamma,\omega\right)  \left(  r-p\right)  r\mathrm{d}r
\]
which belong to $C^{2}$. $\blacktriangleright$

\section{Weighted integral transform and reconstruction}

Let $X$, $\Sigma$ be smooth\ manifolds of dimension $n>1$ and $Z\subset
X\times\Sigma$ be a hypersurface. Let $\pi_{X}:Z\rightarrow X,\ \pi_{\Sigma
}:Z\rightarrow\Sigma$ be the natural projections. A family of hypersurfaces
$Z\left(  \sigma\right)  =\pi_{\Sigma}^{-1}\left(  \sigma\right)  $ in $X$ is
defined as well as the family of hypersurfaces $Z\left(  x\right)  =\pi
_{X}^{-1}\left(  x\right)  $ in $\Sigma$. Suppose that there exists a real
function $\Phi\in C^{2}\left(  X\times\Sigma\right)  $ (called generating
function) such that$\ Z=\Phi^{-1}\left(  0\right)  $ and $\mathrm{d}\Phi\neq0$
in $Z\mathrm{.}$ Let $T^{\ast}\left(  X\right)  $ be the cotangent bundle of
$X$ and $T^{\ast}\left(  X\right)  /\mathbb{R}_{+}$ be the bundle of rays in
$T^{\ast}\left(  X\right)  $. We suppose that

(\textbf{i})\ The\textit{\ }map\textit{\ }$\pi^{\ast}:Z\times\mathbb{R}%
_{+}\rightarrow T^{\ast}\left(  X\right)  $\textit{\ }is a\textit{\ }%
diffeomorphism\textit{,} where $\pi^{\ast}\left(  x,\sigma;t\right)  =\left(
x,t\mathrm{\mathrm{d}}_{x}\Phi\left(  x,\sigma\right)  \right)  ,\ t\in
\mathbb{R}_{+}.$ This condition implies that $\pi_{X}$ has rank $n$ and
$Z\left(  \sigma\right)  $ is a smooth hypersurface in $X$ for any $\sigma
\in\Sigma.$ The function$\ \det J\left(  \Phi\right)  $ vanishes if where
$\Pi^{\ast}$ is not a local diffeomorphism and vice versa.

\textbf{Definition. }We say that points $x\neq y\in X$ are conjugate with
respect to $\Phi$\textit{,} if $\Phi\left(  x,\sigma\right)  =\Phi\left(
y,\sigma\right)  $ and \textrm{d}$_{\sigma}\Phi\left(  x,\sigma\right)
=\mathrm{d}_{\sigma}\Phi\left(  y,\sigma\right)  $ for some $\sigma\in\Sigma.$
We call a generating function $\Phi$ \textit{regular,} if the projection
$\pi:Z\rightarrow X$ is proper, $\Phi$ satisfies the conditions (\textbf{i})
and (\textbf{ii}): there are no conjugate points.

Let $\mathrm{d}X$ be a volume form in $X$ and $b=b\left(  x,\sigma\right)  $
be a continuous function in $Z.$ For any bounded arbitrary function $f$ in $X$
with compact support, the integral%
\[
\mathrm{M}_{b}f\left(  \sigma\right)  \doteqdot\int\delta\left(  \Phi\left(
x,\sigma\right)  \right)  f\left(  x\right)  b\left(  x,\sigma\right)
\mathrm{d}X=\lim_{\varepsilon\rightarrow0}\frac{1}{2\varepsilon}%
\int_{\left\vert \Phi\left(  \cdot,\sigma\right)  \right\vert \leq\varepsilon
}f\left(  x\right)  b\left(  x,\sigma\right)  \mathrm{d}X
\]
converges for any $\sigma\in\Sigma.$ We suppose now that $\mathrm{d}X$ is the
Riemannian volume form of a Riemannian metric $\mathrm{g}$ in $X$. We have:
\begin{equation}
\mathrm{M}_{b}f\left(  \sigma\right)  =\int_{Z\left(  \sigma\right)  }%
\frac{f\left(  x\right)  b\left(  x,\sigma\right)  }{\left\vert \mathrm{d}%
_{x}\Phi\left(  x,\sigma\right)  \right\vert _{\mathrm{g}}}\mathrm{d}%
_{\mathrm{g}}S,\label{1}%
\end{equation}
where $\mathrm{d}_{\mathrm{g}}S$ is the Riemannian volume form in a
hypersurface in $X.$ Here $Z\left(  \sigma\right)  =\left\{  x\in
X,\Phi\left(  x,\sigma\right)  =0\right\}  $ and $\left\vert \cdot\right\vert
_{\mathrm{g}}$ means the Riemannian norm of a covector. Let $\beta$ be a
locally bounded function in $X\times\Sigma.$ The singular integral%
\begin{equation}
\Theta_{b,\beta}\left(  x,y\right)  =\left(  -1\right)  ^{n}\int_{Z\left(
y\right)  }\frac{b\left(  y,\sigma\right)  \beta\left(  x,\sigma\right)
}{\left(  \Phi\left(  x,\sigma\right)  -i0\right)  ^{n}}\frac{\mathrm{d}%
\Sigma}{\mathrm{d}_{\sigma}\Phi\left(  y,\sigma\right)  }\label{7}%
\end{equation}
is well defined for any $x,y\in X,$ $y\neq x$ since $\mathrm{d}_{\sigma}%
\Phi\left(  x,\sigma\right)  \neq0$ in $Z\left(  y\right)  \doteqdot\left\{
\sigma\in\Sigma;\Phi\left(  y,\sigma\right)  =0\right\}  $ due to (\textbf{ii}).

\begin{theorem}
Suppose that a regular generating function $\Phi$ satisfies condition
(\textbf{iii}):
\[
\operatorname{Re}i^{n}\Theta_{b,\beta}\left(  x,y\right)  =0\ \text{for any
}x,y\in X,\ x\neq y.
\]
Then an arbitrary function $f\in L_{2\mathrm{comp}}\left(  X\right)  $ can be
reconstructed from data of $\mathrm{M}_{b}f$ by%
\begin{equation}
D_{b,\beta}f=\mathrm{N}_{\beta}\mathrm{M}_{b}f+\Theta_{b,\beta}f\label{12}%
\end{equation}
where $\Theta_{b,\beta}$ is an operator with the kernel $\Theta_{b,\beta
}\left(  x,y\right)  $ and for even $n$,
\[
\mathrm{N}_{\beta}g\left(  x\right)  =\frac{\left(  n-1\right)  !}%
{\mathrm{j}^{n}}\int_{Z\left(  x\right)  }\frac{g\left(  \sigma\right)
\beta\left(  x,\sigma\right)  }{\Phi\left(  x,\sigma\right)  ^{n}}%
\frac{\mathrm{d}\Sigma}{\mathrm{d}_{\sigma}\Phi},
\]
for odd $n,$%
\[
\mathrm{N}_{\beta}g\left(  x\right)  =\frac{1}{2\mathrm{j}^{n-1}}%
\int_{Z\left(  x\right)  }\delta^{\left(  n-1\right)  }\left(  \Phi\left(
x,\sigma\right)  \right)  g\left(  \sigma\right)  \beta\left(  x,\sigma
\right)  \frac{\mathrm{d}\Sigma}{\mathrm{d}_{\sigma}\Phi}%
\]
is a bounded operator $H_{\mathrm{comp}}^{\left(  n-1\right)  /2}\left(
\Sigma\right)  \rightarrow L_{2\mathrm{loc}}\left(  X\right)  ;$ and%
\begin{equation}
D_{b,\beta}\left(  x\right)  =\frac{1}{\left\vert \mathrm{S}^{n-1}\right\vert
}\int_{Z\left(  x\right)  }\frac{b\left(  x,\sigma\right)  \beta\left(
x,\sigma\right)  }{\left\vert \mathrm{d}_{x}\Phi\left(  x,\sigma\right)
\right\vert _{\mathrm{g}}^{n}}\frac{\mathrm{d}\Sigma}{\mathrm{d}_{\sigma}\Phi
}.\label{6}%
\end{equation}

\end{theorem}

A proof is given in \cite{Pa} for the case $b=\beta=1.$ The general case can
be obtained in the same lines.

\end{document}